\documentclass[reprint,prx,twocolumn,twoside,aps,longbibliography]{revtex4-1}

\usepackage{graphicx}

\usepackage{amsmath,amssymb}
\usepackage{bm}
\usepackage{natbib}
\usepackage{color}
\usepackage{soul,xcolor}

\setstcolor{red}

\begin{document}

\title{The role of geometry in the superfluid flow of nonlocal photon fluids} 
\author{David Vocke$^{1\dagger}$, Kali Wilson$^{1\dagger}$, Francesco Marino$^{2,3}$, Iacopo Carusotto$^4$, Ewan M. Wright$^{5,1}$, Thomas Roger$^{1}$, Brian P. Anderson$^{5}$, Patrik {\"O}hberg$^{1}$, Daniele Faccio$^{1}$}
\email{d.faccio@hw.ac.uk. $\dagger$  these authors contributed equally\\}
\affiliation{$^1$Institute of Photonics and Quantum Sciences, Heriot-Watt University, Edinburgh EH14 4AS, UK} 
\affiliation{$^2$CNR-Istituto Nazionale di Ottica, L.go E. Fermi 6, I-50125 Firenze, Italy}
\affiliation{$^3$INFN, Sez. di Firenze, Via Sansone 1, I-50019 Sesto Fiorentino (FI), Italy}
\affiliation{$^4$INO-CNR BEC Center and Dipartimento di Fisica, Universit\`a di Trento, I-38123 Povo, Italy}
\affiliation{$^5$College of Optical Sciences, University of Arizona, Tucson, AZ, USA}

\begin{abstract}
Recent work has unveiled a new class of optical systems that can exhibit the characteristic features of superfluidity.  One such system relies on the repulsive photon-photon interaction that is mediated by a thermal optical nonlinearity and is therefore inherently nonlocal due to thermal diffusion. Here we investigate  how such a nonlocal interaction, which at a first inspection would not be expected to lead to superfluid behavior, may be tailored by acting upon the geometry of the photon fluid itself. Our models and measurements show that restricting the laser profile and hence the photon fluid to a strongly elliptical geometry modifies thermal diffusion along the major beam axis and reduces the effective nonlocal interaction length by two orders of magnitude. This in turn enables the system to display a characteristic trait of superfluid flow: the nucleation of quantized vortices in the flow past an extended physical obstacle. These results are general and apply to other nonlocal fluids, such as dipolar Bose-Einstein condensates, and show that ``thermal'' photon superfluids provide an exciting and novel experimental environment for probing the nature of superfluidity, with applications to the study of quantum turbulence and analogue gravity.	
\end{abstract}

\maketitle

Over the past decade, theoretical and experimental studies of photon fluids have opened new routes to realizing quantum many-body systems. In the most general case, a photon fluid is created by propagating a laser beam  through a defocusing nonlinear medium such that the photons in the beam act as a gas of weakly interacting particles \cite{Carusotto2013, Carusotto2014,Larre2015, Chiao1999,Weiss1996, Frisch1992}. As in atomic many-body systems, the collective photon behavior can be described by a Gross-Pitaevskii equation \cite{Pethick2008}, where the electric field plays the role of the order parameter, a macroscopic wave function with a clear resemblance to dilute-gas Bose-Einstein condensates (BECs) and superfluid helium.  Much research with photon fluids has therefore centered on exploring characteristics of superfluidity, such as the nucleation of vortices of quantized circulation in the flow past an obstacle for sufficiently high flow speeds \cite{Tilley1990}. \\
\indent Significant progress on this front has been made in the field of {\emph{polariton}} fluids, strongly coupled exciton-polaritons in semiconductor microcavities, that have shown superfluid characteristics, such as the frictionless flow around an obstacle \cite{Amo2009}, and shedding of solitons \cite{Amo2011} and quantized vortices \cite{Sanvitto2011, Nardin2011}. Differently from these driven-dissipative systems with a {\emph{local}} nonlinearity, signatures of superfluid behavior in the dispersion relation have also been observed for room-temperature, {\emph{nonlocal}} photon fluids in a propagating geometry \cite{Vocke2015}. In such a system, heating induced by a laser beam leads to a decrease in the refractive index of the propagation medium and a consequent defocusing effect for the laser beam, resulting in an effective repulsive photon-photon interaction established by the medium. For a monochromatic beam, the transverse beam profile acts to define the geometry of the two-dimensional (2D) fluid and is fully described by hydrodynamic equations, where the propagation direction maps  on to an effective time coordinate \cite{Carusotto2014, Larre2015}.  Due to heat diffusion, this interaction is inherently nonlocal, i.e., the photon density at one point in space influences the interactions at a distant point. Photon-photon interactions are therefore effectively spread out over distances of the order of the characteristic thermal diffusion (or nonlocal) length. While solitons and vortex solitons have previously been reported in these systems \cite{Swartzlander1992,Conti2009a}, the nonlocal aspect of the interaction has only  recently attracted attention \cite{Ghofraniha2012}, enabling studies of novel shock front and turbulence dynamics \cite{Wan2006,Ghofraniha2007,Picozzi2015}. {The main effect of the fluid nonlocality in the context of superfluidity is a reduction and even cancellation of the effect of interactions on intermediate length scales and a restoration of the standard single-particle dispersion relation for excitations in the medium at much lower wavevectors than the inverse healing length. While one can expect that this effect may result in the inhibition of superfluid behavior in generic nonlocal superfluids such as dipolar atomic BECs, the interplay of nonlocality and superfluidity has not yet been completely explored in the literature.}\\
\indent Here we analyze in detail the nature of a propagating laser beam's thermal photon-photon interactions mediated by a dilute dispersion of graphene nanoflakes in methanol.  Our primary result is that with a suitable configuration of the photon fluid's geometry, as determined by the laser beam's transverse profile, nonlocal interactions can exist simultaneously with characteristics of superfluidity, as evidenced by the nucleation of quantized vortices in the photon fluid's flow past an obstacle.  Analytically, we start with a model that accounts for the boundary conditions of the system. By introducing a distributed loss term in the heat diffusion equation, we recover and thus justify the Lorentzian response function that has been used in many previous works \cite{Bar-Ad2013,Vocke2015}. We then show that, for fixed boundary conditions, changing the geometry of the photon fluid, e.g. by using a highly elliptical rather than circular beam, enables us to control and reduce the effective nonlocal interaction length by two orders of magnitude. This choice of beam geometry has the primary effect of restoring locality to a sufficient extent so as to restore  superfluid behavior. With such an elliptical beam, we experimentally demonstrate evidence of superfluidity in nonlocal photon fluids revealed by the nucleation of quantized vortices in the wake of an extended obstacle.\\ 

\begin{figure}[t!]
\includegraphics[width=0.48\textwidth]{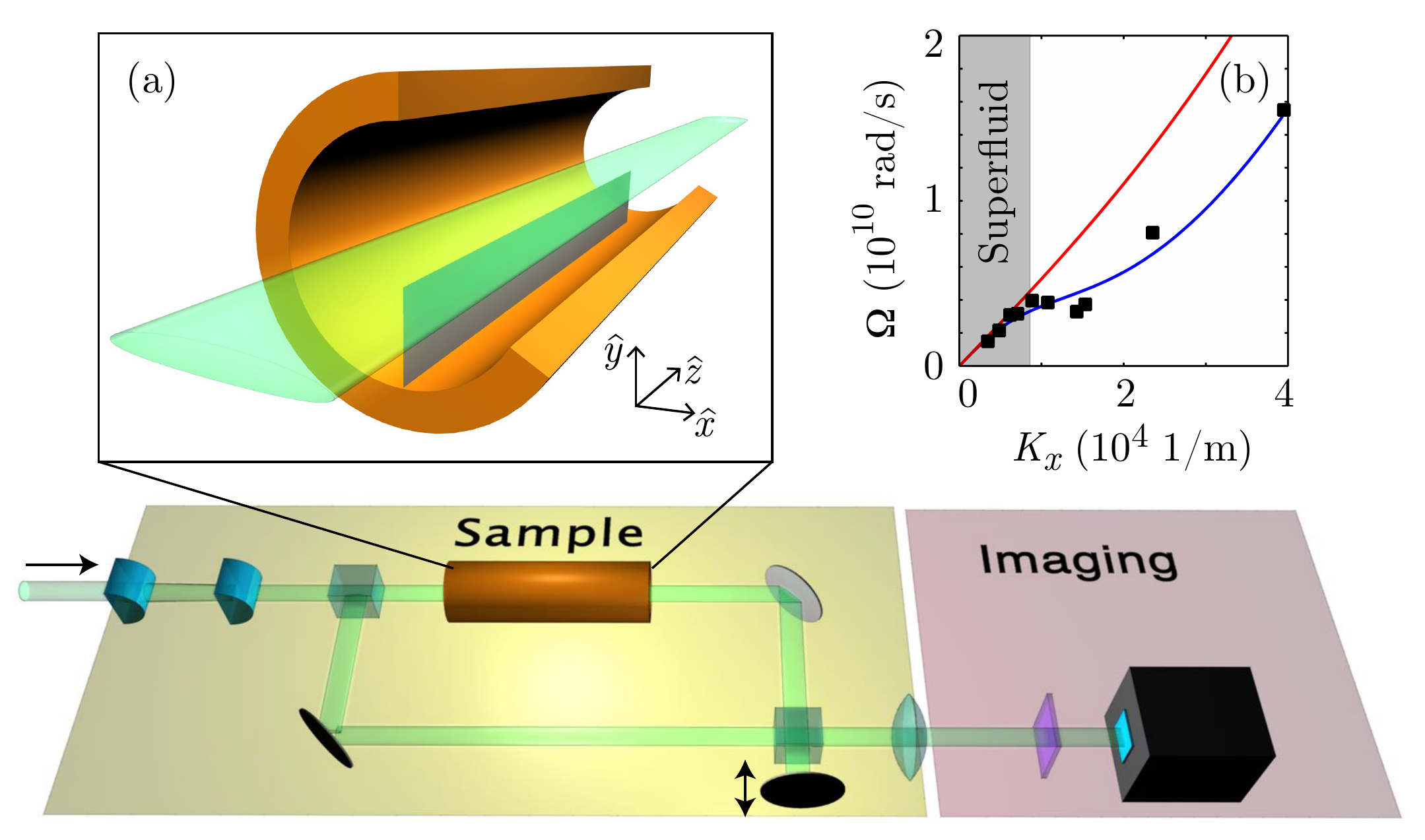}
 \caption{(a) Experimental setup - A 532-nm laser beam with an elliptical spatial beam profile is launched into a sample filled with a methanol-graphene solution. A knife blade is mounted inside the sample and slightly tilted with respect to the propagation direction at an angle $\theta$ that determines the flow of the photon fluid [inset (a)]. Inset (b) shows the measured nonlocal dispersion relation of the photon fluid from Ref.~\cite{Vocke2015}. Solid red line - Bogoliubov dispersion relation, Eq.~(\ref{eq:bogodr_nl}) with a local nonlinearity $|\Delta n| = 7.6\times 10^{-6}$. Solid blue line -  nonlocal dispersion relation with $\hat{R}_{1D}(K_x) = [1 + \bar{\sigma}^2K_x^2]^{-1}$, and $\bar{\sigma} =110$ $\mu\mathrm{m}$ and  $|\Delta n|=7.6\times 10^{-6}$ determined from the best fit to the experimental data (black circles). }
  \label{fig:fig1}
\end{figure}

{\emph{Photon fluids:}} Although the findings of this work are general and apply to any nonlocal superfluid, we derive the main conclusions by focusing our attention on the specific case of a thermal photon fluid where the photon-photon interaction is mediated by heating induced by the local intensity of a propagating laser beam.  Details of the system we are analyzing can be found in Ref.~\cite{Vocke2015} and are summarized in Fig.~\ref{fig:fig1}, which shows the experimental layout used in the experiments described below. The essential component in the system is the actual nonlinear medium, a cylindrical cell with a radius $W=1$ cm and length $L = 18$ cm, filled with methanol  (whose negative thermo-optic coefficient  provides the repulsive photon-photon interaction) and a low concentration of graphene nanoflakes (on the order of $23 \times 10^{-6} ~\mathrm{g}/\mathrm{cm}^{3}$), providing a weak absorption of the input beam needed to enhance the nonlinearity.\\
\indent We briefly recall the main features and equations describing photon fluids. The nonlinear propagation of a laser beam of electric field amplitude $E$ in a defocusing medium can be described within the paraxial approximation \cite{Boyd2008} in terms of a nonlinear Schr\"odinger equation,
\begin{equation} \label{eq:nlse}
\frac{\partial E}{\partial z} = \frac{i}{2k}\nabla^2_\perp E + i(k/n_0) \Delta n E - \frac{\alpha}{2} E.
\end{equation}
Here, $k=2\pi n_0/\lambda$ is the optical wavenumber, $\lambda$ is the vacuum wavelength of the light, and $n_0$ is the linear refractive index. In the limit of negligible absorption $\alpha$, and a local interaction such that $\Delta n=n_2|E|^2$, with $n_2$ the negative nonlinear refractive index,  Eq.~\eqref{eq:nlse} is formally identical to the Gross-Pitaevskii equation for dilute boson gases with a repulsive interatomic interaction \cite{Pethick2008}. By considering the propagation direction $z$ as time coordinate $t=zn_0/c$, where $c$ is the speed of light in vacuum, and the electric field $E(\mathbf{r},t)=\sqrt{\rho(\mathbf{r},t)}\exp{(i\phi(\mathbf{r},t))}$ as a function of photon fluid density $\rho = |E|^2$  (with units of $\mathrm{V}^2/\mathrm{m}^2$) and phase $\phi$, one arrives at a set of hydrodynamic equations over transverse spatial coordinate $\boldsymbol{r} = (x,y)$ that describes the laser beam as a 2+1-dimensional quantum fluid of light \cite{Chiao1999, Carusotto2014, Frisch1992},
\begin{eqnarray}
	\partial_{t}\rho+\boldsymbol{\nabla}_\perp \cdot (\rho \boldsymbol{v})=0 \label{eq:conti} \\
	\partial_{t}\psi+\frac{1}{2}v^2+\frac{c^2n_2}{n_0^3}\rho-\frac{c^2}{2k^2n_0^2}\frac{\nabla^2_\perp \sqrt{\rho}}		{\sqrt{\rho}}=0.
	\label{eq:euler}	
\end{eqnarray}
The transverse gradient of the phase of the  beam determines the transverse fluid flow velocity $\boldsymbol{v}=(c/kn_0)\boldsymbol{\nabla}_\perp \phi=\boldsymbol{\nabla}_\perp \psi$ and the speed of long-wavelength sonic waves is given by $c_s=\sqrt{c^2\vert n_2\vert\rho/n_0^3}$. Here $\psi = (c/kn_0)\phi$ is proportional to the phase of the wave function, and by taking the transverse gradient, Eq.~\eqref{eq:euler} may be recast as the equation of motion for $\boldsymbol{v}$ in direct analogy to the hydrodynamic Euler equation for dilute-gas BECs \cite{Pethick2008}. Transverse wave perturbations propagating on an intense plane-wave beam obey the well known Bogoliubov dispersion relation \cite{Chiao1999, Carusotto2014, Marino2008}.
\begin{equation}
	(\Omega-vK)^2=\frac{c^2\vert n_2\vert
	\rho}{n_0^3}\hat{R}(K,n_0\Omega/c)K^2+\frac{c^2}{4k^2n_0^2}K^4,
	\label{eq:bogodr_nl}
\end{equation} 
relating frequency $\Omega$ and wavenumber $K=\sqrt{K_x^2+K_y^2}$ of phonon excitations.
Here $\hat{R}$ is the Fourier transform of the medium response function. In the local case, $\hat{R}(K,n_0\Omega/c)=1$ but in the nonlocal case $\hat{R}(K,n_0\Omega/c)$ will take on a more complicated functional form that will ultimately determine the extent of superfluid behavior observable in the photon fluid. It is precisely the nature of this response function that we investigate in more detail in the following.\\

\begin{figure}[t!]
	\includegraphics[width=7cm]{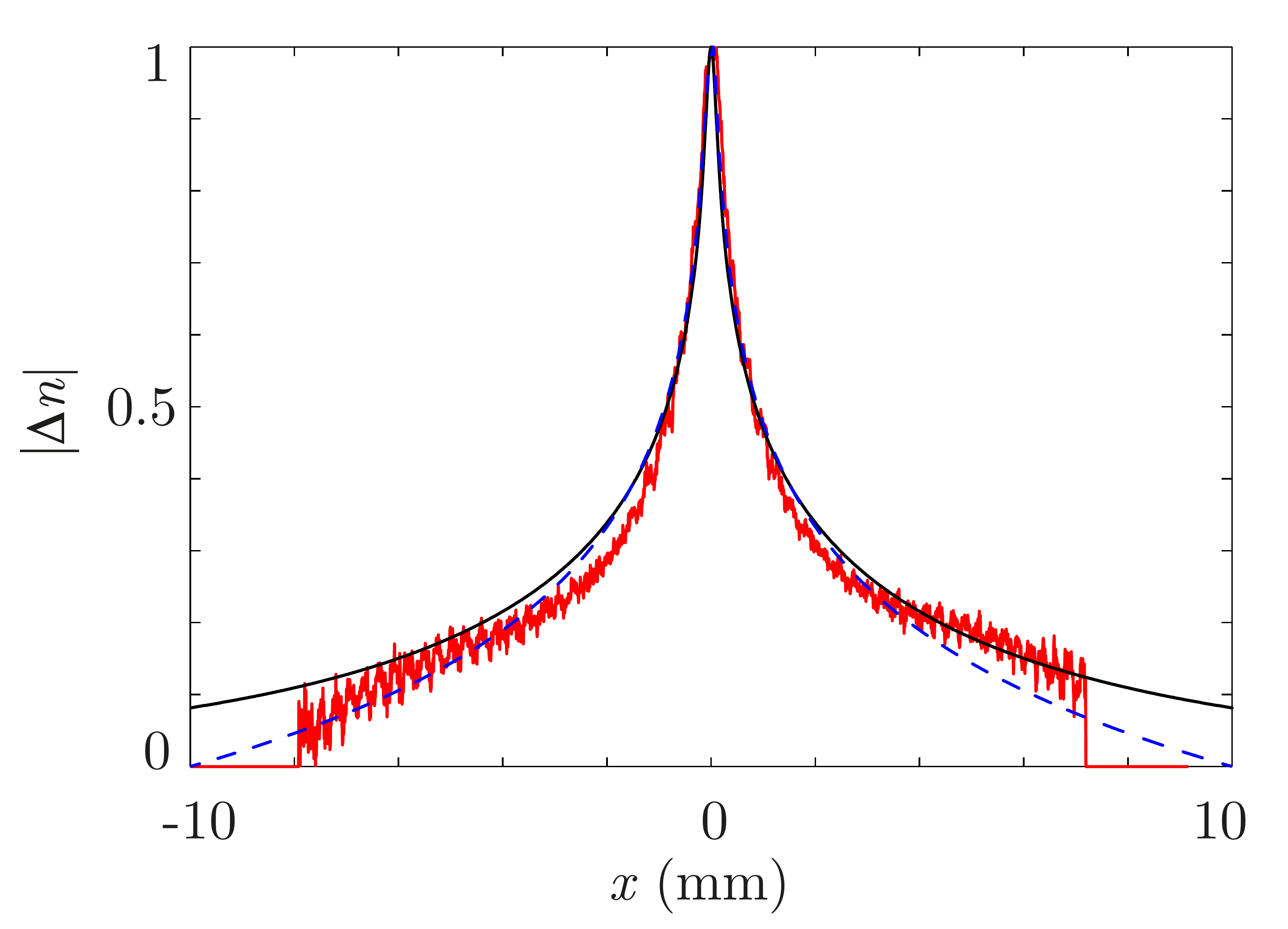}
 	\caption{Cross section of the real-space, thermally-induced refractive index variation, $|\Delta n(x,y=0)|$. Red curve: experimental measurement using a methanol-graphene medium with $L = 12$ cm, $W = 1$ cm, $\alpha \sim 0.013 ~\mathrm{cm}^{-1}$, similar to that used in the experiments in Fig.~\ref{fig:flowobstacle}. Dashed blue curve: the numerical solution to Eq.~\eqref{eq:heat1} with a physical circular boundary at $W = 1$ cm. Black curve: the analytical solution from the DLM model with $\sigma=W=1$ cm. All curves are normalized to one.}
   	\label{fig:DLM}
\end{figure}

{\emph{Thermal, nonlocal nonlinearity:}} We now consider in detail the case of {\emph{nonlocal}} photon-photon interactions. { The cell used in our experiments is sufficiently long and the characteristic frequency of the propagation evolution (Eq.~(\ref{eq:nlse})) is sufficiently slow that we may neglect any variations of the heat diffusion profile along $z$.} Under these assumptions, the nonlocal change in the refractive index $\Delta n(x, y)$, at a given propagation plane along $z$, can then be found by convolving the material response function $R$ with the source beam intensity profile $I(x, y)$ on that plane.  We assume a beam that occupies the central region of the fluid, far from the system's physical boundaries, so that the system is approximately shift-invariant with 
\begin{equation}
\Delta n(x,y) = \gamma \int dx' dy' ~R(x-x^\prime,y-y^\prime) I(x^\prime, y^\prime) ,
\end{equation}
with the constant $\gamma$ accounting for the properties of the material, and $ \int dx^\prime dy^\prime R(x-x^\prime,y-y^\prime) = 1$.
For a thermal nonlinearity such as that provided by the methanol-graphene system, the temperature dependent change in the index of refraction is $\Delta n = \beta \Delta T$ where $\beta = dn/dT$. The spatial profile of $\Delta n(x,y)$ is then determined by the 2D steady-state heat equation,
\begin{equation}\label{eq:heat1}
\nabla^2_\perp(\Delta T) = - \frac{\alpha}{\kappa} I(x,y),
\end{equation}
where $\kappa$ is the thermal conductivity.\\
\indent Thus the response function is, to within a constant, the Green's function for heat diffusion in the medium, and obeys
\begin{equation}\label{Req1}
\nabla_\perp^2 R(x-x^\prime,y-y^\prime) =-\left ({\alpha\beta\over\kappa\gamma}\right ) \delta(x-x^\prime,y-y^\prime).
\end{equation}
We can find a numerical solution for $R$ by using a narrow Gaussian source beam $I(x,y) = I_0 e^{-2(x^2/w_x^2 + y^2/w_y^2)}$ to approximate the delta function, and then numerically solving Eq.~\eqref{eq:heat1} with the condition $\Delta T(x,y) = 0$ for $\sqrt{x^2+y^2}=W$.  Our choice of boundary condition assumes that, due to efficient heat transport to the surrounding environment at the boundaries, there is no change in temperature at the boundary itself. \\
\indent Alternatively, we can approximate the effect of the boundary by incorporating a distributed loss term $-\Delta T / \sigma^2$ into the steady-state heat equation,
\begin{equation}
\nabla^2_\perp(\Delta T) - \frac{\Delta T}{\sigma^2} = - \frac{\alpha}{\kappa} I(x,y).
\label{heat}
\end{equation} 
With the addition of this distributed loss term we obtain a Lorentzian $K$-space response function,
\begin{equation}
\hat{R}(K_x,K_y) = \frac{1}{1 + \sigma^2 (K_x^2 + K_y^2)},
\label{response}
\end{equation}
where the nonlocal length $\sigma = W$ is fixed by the system boundary and determines the spatial-frequency cutoff of the response function. In $K$-space the nonlocal change in the index of refraction is $\Delta \hat{n}(K_x,K_y) = \hat{\gamma} \hat{R}(K_x,K_y) \hat{I}(K_x,K_y)$, with $\hat{\gamma} = \alpha \beta \sigma^2/ \kappa$ the effective nonlinear coefficient. The response function defined in Eq.~\eqref{response}, which we derive from the distributed loss model (DLM), has been taken as an a priori assumption in previous works \cite{Bar-Ad2013,Vocke2015}. The DLM solution proves to be a valid  analytical approximation to the exact numerical solution of Eq.~\eqref{eq:heat1}.
Indeed, the corresponding DLM real-space response function 
\begin{equation}
R(\boldsymbol{r}-\boldsymbol{r}^\prime) = \frac{1}{2\pi\sigma^2} K_0\left(\frac{\boldsymbol{r}-\boldsymbol{r}^\prime}{\sigma}\right),
\end{equation}
where $\boldsymbol{r} = (x,y)$ and $K_0$ is the zeroth-order modified Bessel function of the second kind, is shown in Fig.~\eqref{fig:DLM} to provide quantitative agreement with the numerical solution to the heat equation assuming a narrow Gaussian source beam. In the same figure we also show the measured  $|\Delta n(x,y=0)|$ (red curve), obtained using an interferometric method similar to that used by Minovich \emph{et al.} \cite{Minovich2007}. Summarizing this first result, the experimental measurement of the response function indicates that  the DLM provides a good approximation to both the measured $|\Delta n(x,y)|$ and to the exact numerical solution to Eq.~\eqref{eq:heat1}.\\  
\indent We next turn to the analytical DLM solution to describe how the beam or fluid geometry may be used to control the nonlocal interaction length. We first point out that for a medium with physical boundaries, as in the experiments described below and in Ref.~\cite{Vocke2015}, a cylindrically symmetric beam with $1/e^2$ beam radii $w_x, w_y \sim W$ implies superfluid behavior only for very low wavevectors such that $K_x, K_y < 1/\sigma$ where the response function approaches a local response $\hat{R}(K_x,K_y) = 1$.  However, accessing these wavevectors is experimentally unrealistic as it requires observation and measurement of waves in the transverse plane of the beam that have a wavelength that is of the order of the size of the medium itself.\\
\indent However, the situation is more interesting in the case of an elliptical beam geometry with $w_y << w_x \simeq W$.  The focused beam radius $w_y<<w_x$ leads to a temperature distribution whose width scales with $w_y$ along the $y$-axis.  This focusing along the $y$-axis also leads to enhanced thermal diffusion away from the major axis of the beam aligned with the $x$-axis, as may be seen by applying the approximation ${\partial^2 (\Delta  T)\over\partial y^2} \approx -{\Delta T\over w_y^2}$ in Eq.~\eqref{heat}.  Within this approximation we may use an effective one-dimensional heat equation, resulting in an effective 1D response function  $\hat{R}_{1D}(K_x) = 1/(1+\bar{\sigma}^2K_x^2)$ with the introduction of an effective nonlocal length, $\bar{\sigma} = (\sigma w_y)/ \sqrt{w_y^2 + \sigma^2}$.  The key point here is that focusing along the $y$-axis decreases the length scale for thermal diffusion to $\bar{\sigma} \sim w_y$ along both major and minor beam axes. Thus our elliptical geometry can lead to superfluid behavior for the range $K_x, K_y < 1/w_y$, which is now experimentally accessible. A circular beam with $w_x  = w_y << W$ may seem an option, but is impractical from an experimental standpoint since the spatial extent of the fluid would then be too small to observe superfluid behavior such as vortex nucleation in the flow past an obstacle.\\ 
\indent We arrive at an alternative view of the effective geometry-induced tailoring of the medium response by considering the spatial extent of the focused spot size $w_y$ and the associated $y$-component of the wavevector $K_y = {1/ w_y}={{2\pi} / {\Lambda_y}}$, where $\Lambda_y$ is largest wavelength oscillation supported by the spatial extent of the minor axis of the elliptical photon fluid, such that $\hat R_{1D}(K_x)=\hat{R}(K_x,K_y=1/w_y)$. We may then  determine the $K$-space response function by first numerically solving the full 2D heat equation and taking the Fourier transform to obtain $\hat{R}(K_x,K_y)$.  We found that either integrating the response function over all $K_y > 1/w_y$ or simply taking $\hat{R}_{1D}(K_x) = \hat{R}(K_x, K_y = 1/w_y)$ gives similar functions. In the latter case, the DLM response function may again be rewritten as $\hat{R}_{1D}(K_x) = 1/(1+\bar{\sigma}^2K_x^2)$, and together with Eq.~\eqref{eq:bogodr_nl} provides good agreement with the dispersion relation reported in Ref.~\cite{Vocke2015} (see Fig.~\ref{fig:fig1}). Therefore, the DLM approach explicitly indicates that the highly elliptical geometry of the beam leads to an effective 1D response function with a nonlocal length, which in the limit $w_y << \sigma$ loses all trace of the physical boundary conditions and is essentially determined solely by the beam or fluid geometry,  $\bar{\sigma} \sim w_y$. This provides a clear route to re-establishing and controlling the onset of superfluid behavior in an otherwise highly nonlocal and non-superfluid medium. \\

{\emph{Superfluidity and vortices:}}  In order to verify our findings we performed a series of experiments and numerical simulations with both round and highly elliptical pump beams, aimed at observing a characteristic trait of superfluid flow, namely quantized vortex nucleation in supercritical flows around an obstacle. We first briefly describe the characteristics of quantized vortices in superfluids. The physics of quantized vortices has been of great interest ever since their discovery in  $^4$He \cite{Onsager1949, Feynman1955, Vinen1957} and later in BECs \cite{Matthews1999}, and their inherent dynamics are intensively studied in quantum turbulence \cite{Barenghi2014, White2014}. Unfortunately, due to the relatively short coherence length in $^4$He, the vortex core diameter is of the order of a few \AA ngstr\"oms and is hard to visualize by optical means, although progress has been made using hydrogen tracer particles \cite{Paoletti2008}. In contrast, in BECs and photon fluids, the diluteness of the fluid leads to larger healing lengths, with vortex core diameters on the order of half a micron and tens of microns respectively, allowing for direct observation \cite{Wilson2015, Swartzlander1992, Sanvitto2011, Nardin2011}. Furthermore, in photon fluids straightforward optical interferometry provides easy access to phase information and thus photon fluids provide an accessible platform to study quantized vortex dynamics, where core position, circulation and winding number can be easily identified. \\

\begin{figure}[t!]
	\includegraphics[width=0.48\textwidth]{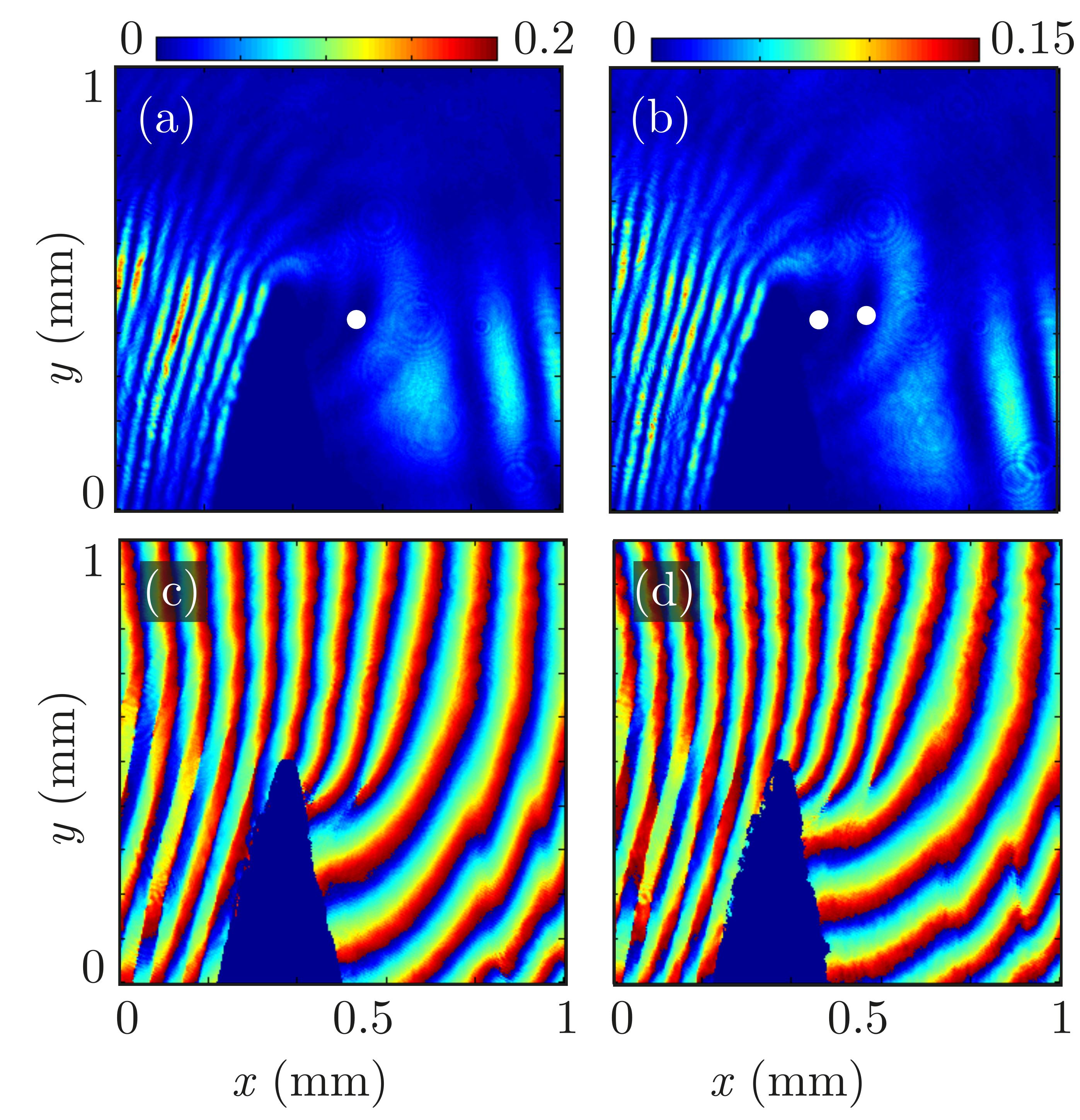}
 	\caption{Experimental measurements of the near-field beam profile at the sample output $I(x,y)/I_0$, normalized by input beam intensity $I_0$. (a) and (b) $I(x,y)/I_0$ showing superfluid instability at high flow speed,  $v_f\sim1.3 \times 10^6$ m/s: (a) $I_0 = 4.0$ W/cm$^2$,  i.e., $\Delta n = - 3.0\times10^{-5}$ and $v_f/c_s\sim1.2$,  and (b) $I_0 = 6.8$ W/cm$^2$,  i.e., $\Delta n = - 5.2\times10^{-5}$ and $v_f/c_s\sim0.9$. White circles indicate the position of the vortex singularities obtained from the corresponding phase diagrams shown in (c) and (d), respectively. }
   	\label{fig:flowobstacle}
\end{figure}

{\emph{Experiments:}} The experimental setup is shown in Fig.~\ref{fig:fig1}. We launch a highly elliptical continuous-wave laser beam with wavelength $\lambda=532$  nm through a cylindrical cell with length $L=18$ cm and radius $W = 1$ cm, filled with a methanol-graphene solution as a thermal nonlinear medium. As shown in inset (a) of Fig.~\ref{fig:fig1}, the beam propagates along $z$, with its minor axis oriented along $y$. Methanol has a negative thermo-optic coefficient of $\beta=-4\times 10^{-4}$ 1/K (providing the repulsive photon-photon interaction) and thermal conductivity $\kappa = 0.2$ W/(m$\cdot$K), while nanometric graphene flakes (on the order of 1.6 mg of graphene in the 60 cm$^3$ volume of methanol) are added in order to increase the absorption coefficient to $\alpha=0.035$ cm$^{-1}$ and ensure sufficient thermo-optic nonlinearity at the input intensities (of the order 4-7 W/cm$^2$) used in the experiments.  The beam is loosely focused onto the sample by a set of cylindrical lenses with a minor axis $1/e^2$ beam radius of $210~\mu$m (the major axis radius is $\sim1$ cm). The sample is placed in one arm of a Mach-Zehnder interferometer. The other arm is used  to probe the spatial phase profile of the sample output plane with a reference beam. To this end, a piezo-controlled delay stage is introduced in order to retrieve the full phase information by a phase-shifting interferometry technique \cite{Kinnstaetter1988}. Finally, the intensity profile of the beam at the cell output is imaged with 4x magnification onto a CCD camera. As an obstacle, a knife blade is immersed in the sample along the beam path [see inset (a), Fig.~\ref{fig:fig1}] so that the beam propagation axis is at a slight angle $\theta$ with respect to the knife axis, thus introducing a relative transverse flow $v_f=(c/n_0)\sin\theta$ along the $x$-axis, determined by the respective angle $\theta$. \\
\indent For excitations with wavenumbers smaller than the inverse of the effective nonlocal length $(K<1/\bar{\sigma})$ the dispersion is a linear function in $K$ ($\Omega\approx c_sK)$. Using the parameters and the technique described in Ref.~\cite{Vocke2015}, we measured the $\Omega(K_x)$ spectrum of elementary excitations and indeed show that these follow the Bogoliubov dispersion with an effective 1D nonlocal length, $\bar{\sigma}= 110~\mu$m and $|\Delta n| =\vert n_2\vert\rho= 7.6 \times 10^{-6}$ determined from the best fit to the experimental data [see Fig.~\ref{fig:fig1}(b)]. Hence there exists a critical velocity $v_c$ for which a fluid with flow speed $v<v_c$ behaves as a superfluid and is not perturbed by small defects inserted in the flow. For local fluids, the critical velocity is determined by the speed of sound. However, as discussed in the Appendix, the nonlocality can significantly reduce the critical velocity. For our experimental parameters, this reduction amounts to approximately a factor of two. Furthermore, in this work we consider an extended impenetrable obstacle such that the local flow speed can become supercritical in its vicinity, even if the asymptotic speed $v_f$ far from the obstacle remains subcritical.\\
\indent The breakdown of superfluidity leads to dissipation and drag forces on the obstacle and consequently, the nucleation of quantized vortices is expected to appear in the local interaction case \cite{Nardin2011, Inouye2001}. Figure~\ref{fig:flowobstacle} shows experimental evidence in our {\emph{nonlocal}} superfluid of the breakdown  of the viscousless flow and subsequent nucleation of quantized vortices, which are considered to be hallmark evidence of superfluid behavior. In detail, Figs.~\ref{fig:flowobstacle}(a) and (b) show the normalized near-field intensity profile at the sample output $I(x,y)/I_0$ for two different input intensities  $I_0 = 4.0$ W/cm$^2$ and $I_0 = 6.8$ W/cm$^2$, respectively (corresponding to two different sound speeds, $v_f/c_s\sim 1.2$ and $v_f/c_s\sim 0.9$) and with $\theta$ chosen to create a flow speed $v_f = 1.3 \times 10^6$ m/s along the positive $x$-axis. 
The intensity profiles show how the rigid obstacle creates a wake in the downstream region, but just as in an ordinary fluid, the light intensity is able to flow around the tip of the obstacle and fill in the shadowed region.
A series of vortices (dark regions in the intensity profile, marked with white dots) are seen to nucleate from close to the tip. One vortex is observed for $v_f/c_s\sim 1.2$ and a second vortex is formed for higher excitation (i.e. sound) speeds, $v_f/c_s\sim 0.9$. This interpretation finds confirmation in the phase profiles shown in Figs.~\ref{fig:flowobstacle}(c) and (d) where the drag on the obstacle is visualized by the bending of the phase pattern around the tip of the obstacle. The nucleation of vortices in the wake of the obstacle is evidenced by the clockwise-circulating phase singularities. We note that increasing the light intensity, and therefore the magnitude of $\Delta n$, both increases the speed of sound of the fluid (decreasing $v_f/c_s$), and also speeds up the effective ``temporal" evolution of the fluid, such that within certain limits increasing $|\Delta n|$ will have the same effect as increasing the propagation length. The observation of two vortices in Fig.~\ref{fig:flowobstacle}(d) is therefore due to the faster sound speed, i.e. faster overall evolution (rather than to an increase in flow speed).\\
\begin{figure}[t!]
\includegraphics[width=0.48\textwidth]{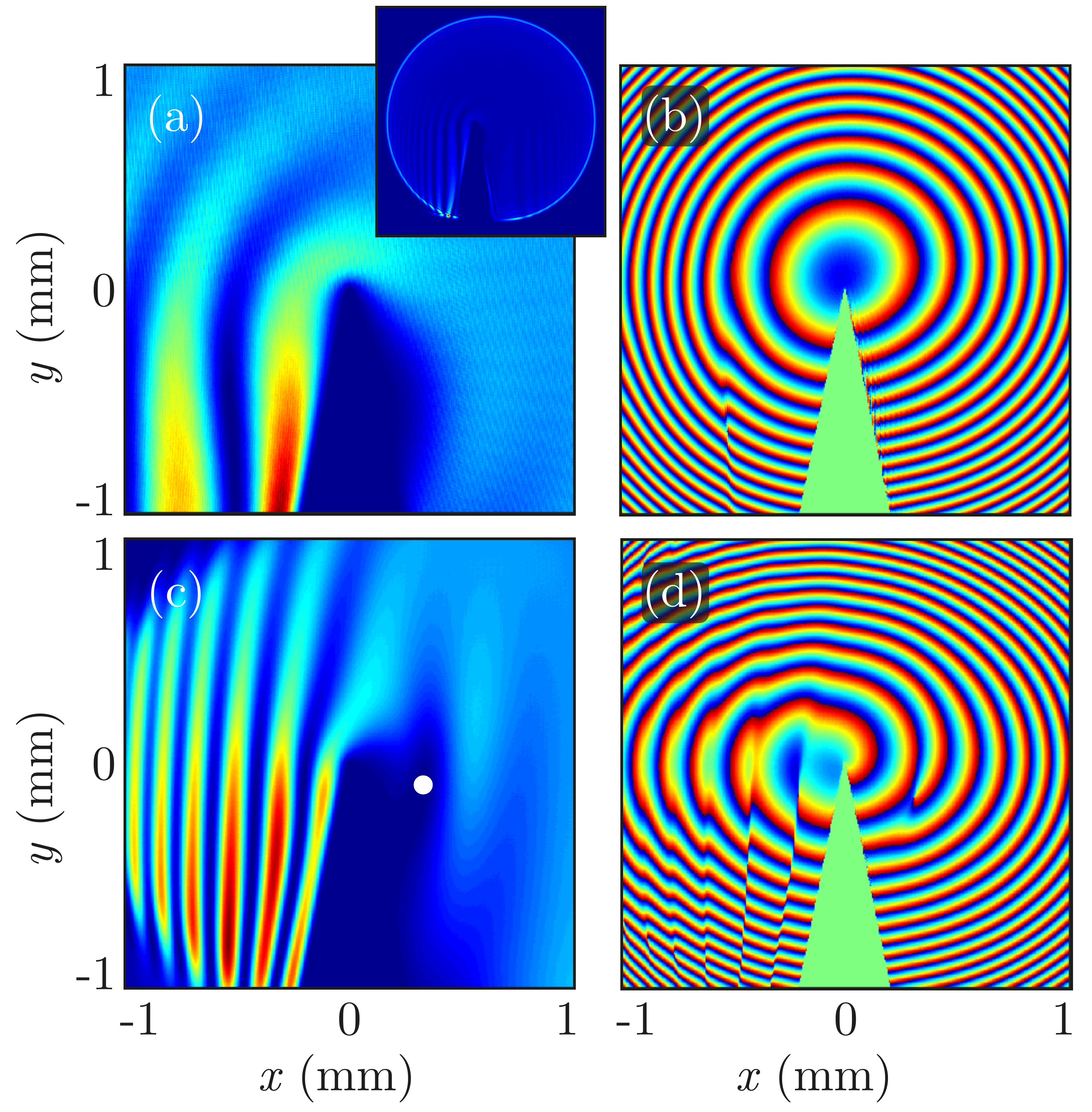}
 \caption{Numerical simulations. Circular input pump beam with 1-mm $1/e^2$ beam radius: (a) intensity profile at the output of the nonlinear medium, zoomed in around the obstacle tip. The inset shows the full beam profile. (b) phase profile for the circular beam (wrapped between $-\pi$ and $+\pi$. The smoothness and absence of singularities indicate a complete absence of vortex nucleation around the obstacle. Elliptical input pump beam ($w_x=1$ mm, $w_y=0.15$ mm): (c) intensity profile with a white circle indicating the position of a vortex nucleated from the obstacle, and (d) the phase profile. A clear phase singularity can be observed confirming the presence of the vortex. Simulations in (a)-(b) and (c)-(d) are performed under identical conditions, aside from the input beam profile, with input intensity $I_0=2.65$ W/cm$^2$, propagation length $L = 18$ cm, fluid flow speed $v_f = 1.0 \times 10^6$ m/s, and nonlocal interaction length $\sigma=1$ cm.}
  \label{numerics}
\end{figure}

{\emph{Numerical simulations:}}  In order to investigate these observations of vortex nucleation in more depth, we performed numerical simulations that were based on split-step propagation of Eq.~\eqref{eq:nlse}, with $\alpha = 0.035 ~\mathrm{cm}^{-1}$ chosen to match the experiment. Nonlocality is described by the DLM 2D Lorentzian response function $\hat{R}(K_x,K_y)$  with $\sigma=1$ cm. The dimensions and shape of the obstacle were chosen so as to match the experiments.  As can be seen in Figs.~\ref{numerics}(a)-(b) (normalized intensity and phase, respectively), when using a round beam no particular features are observed in the  flow around the obstacle. However,  Figs.~\ref{numerics}(c)-(d) show the results for the exact same simulation parameters (in particular, for the same intensity $I_0=2.65$ W/cm$^2$) for an elliptical beam with minor axis $w_y=150$ $\mu$m, and vortex nucleation is clearly observed close to the tip of the obstacle. These numerical results are in good agreement with our experiments and show how by simply controlling the geometry of the beam, or more generally the geometry of the fluid, one may tune the effective nonlocal interaction length $\bar{\sigma}$ and hence re-establish and control superfluid behavior in nonlocal fluids.\\
\begin{figure}[t!]
\includegraphics[width=0.48\textwidth]{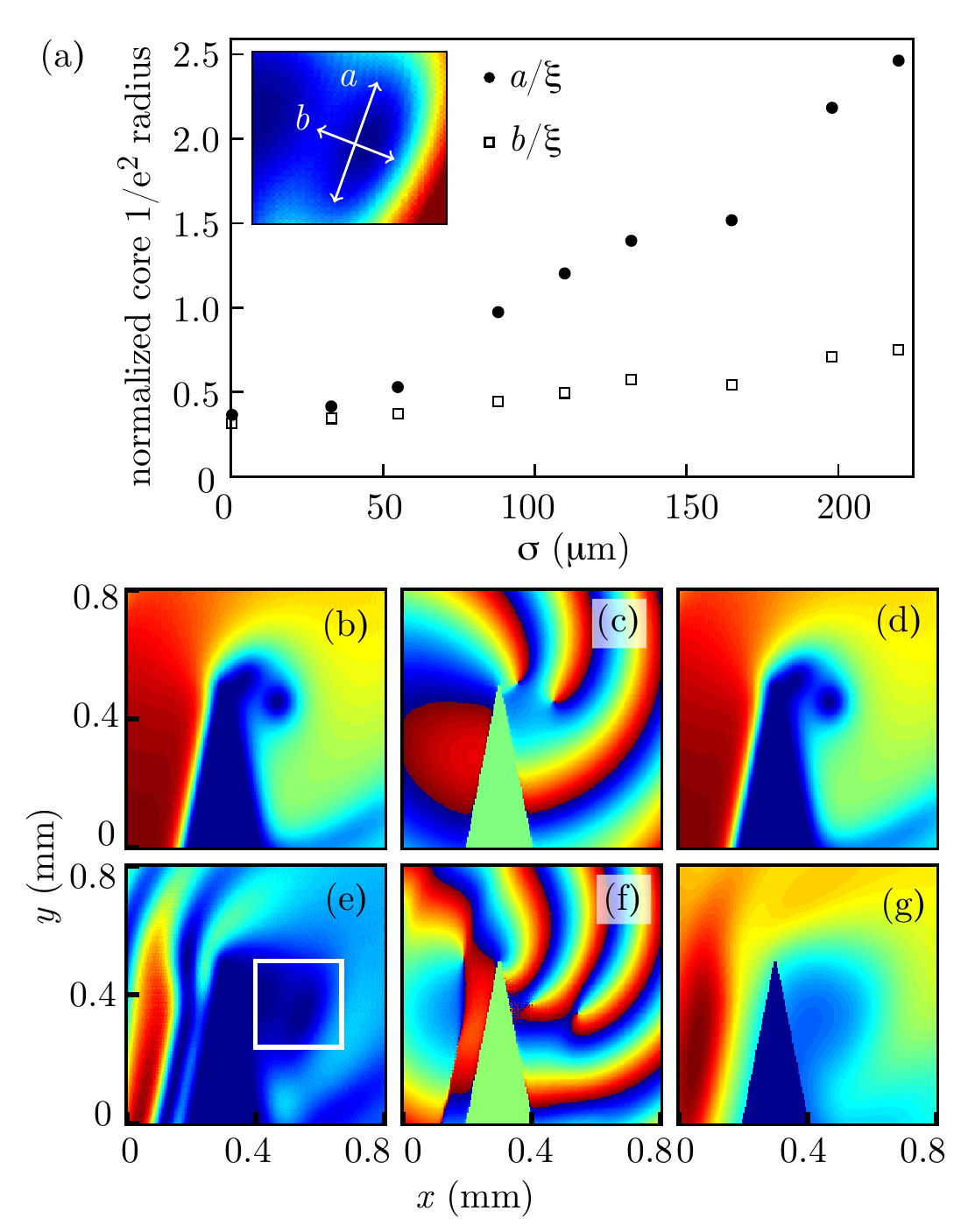}
\caption{Numerical results - (a) Plot of vortex core $1/e^2$ radii, $a$ and $b$, normalized to the local healing length $\xi$, versus $\sigma$. The inset, corresponding to the white region in (e), shows a vortex core with labeled major axis $a$ and minor axis $b$. (b) normalized beam intensity, $I(x,y)/I_0$, (c) phase $\phi(x,y)$ and (d) nonlinear potential, $\Delta n(x,y) / \Delta n$,  for a purely local superfluid after 20 cm of propagation, with  $v_f/c_s = 0.56$, $\Delta n = - 1.2\times10^{-5}$, and $1/e^2$ beam radii $w_\mathrm{x} = w_\mathrm{y}$ = 1 mm. (e), (f) and (g) show the corresponding profiles for a nonlocal superfluid with $\sigma = 110$ $\mu$m, and $\Delta n = - 2.4\times10^{-5}$.  }
\label{fig:sims2}
\end{figure}
\indent As a side note, the vortices nucleated in the presence of nonlocality have an elliptical shape that is observed both in the experiments and numerical simulations described above.  In Fig.~\ref{fig:sims2}(a) we plot the  vortex core $1/e^2$ radii $a$, $b$ (normalized to a local healing length $\xi = \lambda/\sqrt{4n_0 |n_2| \rho}$ so as to account for any local variation in the fluid density associated with the output beam intensity) as a function of increasing nonlocality, $\sigma$.  For each value of the 2D nonlocal interaction length $\sigma$, the input power is chosen such that we observe one clearly defined vortex core, and flow speed is adjusted to maintain a constant $v_f/c_s = 0.56$.  At $\sigma=0~\mu$m (purely local response) the vortex cores are circular. However, as we increase $\sigma$, we observe a continuous increase in the core aspect ratio, $a/b$. 
 Figures~\ref{fig:sims2}(b)-(d) show the normalized intensity $I(x,y)/I_0$, phase $\phi(x,y)$ and effective nonlinear potential energy $\Delta n(x,y) / \Delta n$ profiles, respectively, for a purely local superfluid after 20 cm of propagation. As can be seen in Fig.~\ref{fig:sims2}(b), the nucleated vortices have the expected circular central core region, which is reflected in the effective nonlinear potential energy profile shown in Fig.~\ref{fig:sims2}(d).  Figures~\ref{fig:sims2}(e)-(g) show the corresponding profiles for $\sigma=110~\mu$m, taken as a representative example from the points shown in Fig.~\ref{fig:sims2}(a).   The vortices are now clearly elliptical: in particular, Fig.~\ref{fig:sims2}(g) shows a marked deformation of the effective nonlinear potential energy term so that the spatial extent of the knife blade felt by any point in the fluid is much larger when compared to the local case. We may thus interpret the vortex deformation in terms of a nonlocal smearing of the nonlinear potential along the obstacle edge.  \\
 
\indent {\emph{Conclusions:}}  Large nonlocal interaction lengths will strongly suppress superfluid behavior, effectively removing the photon-photon (or particle-particle) interactions that are at the origin of the linear superfluid dispersion relation. However, by controlling the geometry of the beam or fluid, it is possible to  limit the effective nonlocal interaction length and thus re-establish superfluid behavior. A distributed loss model for heat flow can  quantitatively justify this effect and explain how  we reduced the effective nonlocal length  by two orders of magnitude
in our experiments. Both experiments and numerical simulations with elliptical beams demonstrate the nucleation of quantized vortices from an extended hard obstacle in a flowing photon fluid, which is considered to be the hallmark trait of superfluid behavior. \\
\indent Although it was necessary to consider the full details of our specific system in order to unveil the role of the fluid geometry,  the general conclusions reach beyond thermal photon fluids and apply to any nonlocal superfluid, or, for that matter, to any form of nonlocal interaction. The photon superfluids specifically described here are of relevance in relation to recent proposals for studying horizon and Hawking-like emission in artificial spacetime geometries that mimic Lorentz-invariance based on the superfluid dispersion relation \cite{Marino2008,Fouxon2010,Fleurov2012a,Carusotto2014}, or for cases in which it is the nonlocality itself that is used to mimic long-range gravitational effects \cite{Segev2015}. Thermal photon fluids may also provide an alternative testbed for nonlocal effects observed in dipolar BECs \cite{Conti2014c,Lahaye2009,Pu2008,Lewenstein2003}.  

\section{Acknowledgements}
D.F. acknowledges financial support from the European Research Council under the European Unions Seventh Framework Programme (FP/2007–2013)/ERC GA 306559 and EPSRC (UK, Grant EP/J00443X/1). I. C. acknowledges financial support by the ERC through the QGBE grant, by the EU-FET Proactive grant AQuS, Project No. 640800, and by the Autonomous Province of Trento, partly through the SiQuro project ("On Silicon Chip Quantum Optics for Quantum Computing and Secure Communications").

\section{Appendix}

Here we provide further details on the effect of nonlocal interactions on the Landau critical velocity.
The starting point of this discussion is the Bogoliubov dispersion of collective excitations in such a {\em nonlocal fluid} when this is at rest. After simple manipulation of eq.(4) of the Main Text, this can be written in the more usual form~\cite{Pomeau}:
\begin{equation}
 \hbar\Omega(k)=\sqrt{\frac{\hbar^2k^2}{2m}\left(\frac{\hbar^2k^2}{2m} + \frac{2 g \rho}{1+k^2\sigma_L^2}\right)}
\end{equation}
that allows for transparent analytical manipulations and easier connection to the many-body literature. With respect to the usual case of a local fluid~\cite{LPSS_book,RMP}, the nonlocality enters via the denominator of the interaction term that suppresses interactions at large $k$.
For a propagating fluid of light, the frequency $\Omega$ corresponds to the wavevector $k_z$ in the propagation direction expressed in temporal units, $\Omega=k_z c/n_0$, and $k$ is the transverse wavevector (note the different convention from~\cite{IC-PR}). 

In terms of the optical parameters of the optical medium,  
\begin{equation}
m=\frac{ \hbar n_0 k_0}{c}
\end{equation}
is the effective mass of the photons, $\rho=|E|^2$ is the fluid density,
\begin{equation}
 g=\frac{\hbar c k_0}{n_0^2}\,\chi^{(3)}
 \end{equation}
is the interaction constant, and the nonlocality is again modeled with a Lorentzian function of size $\sigma_L$. Here, $n_0$ is the background refractive index and $k_0=2\pi n_0/\lambda$ the free wavevector in the medium, and $\chi^{(3)}$ is the optical nonlinearity. The units of this latter are such that $\Delta n=\chi^{(3)}\rho$ is the nonlinear refractive index shift.

\begin{figure}[htbp]
\includegraphics[width=0.9\columnwidth]{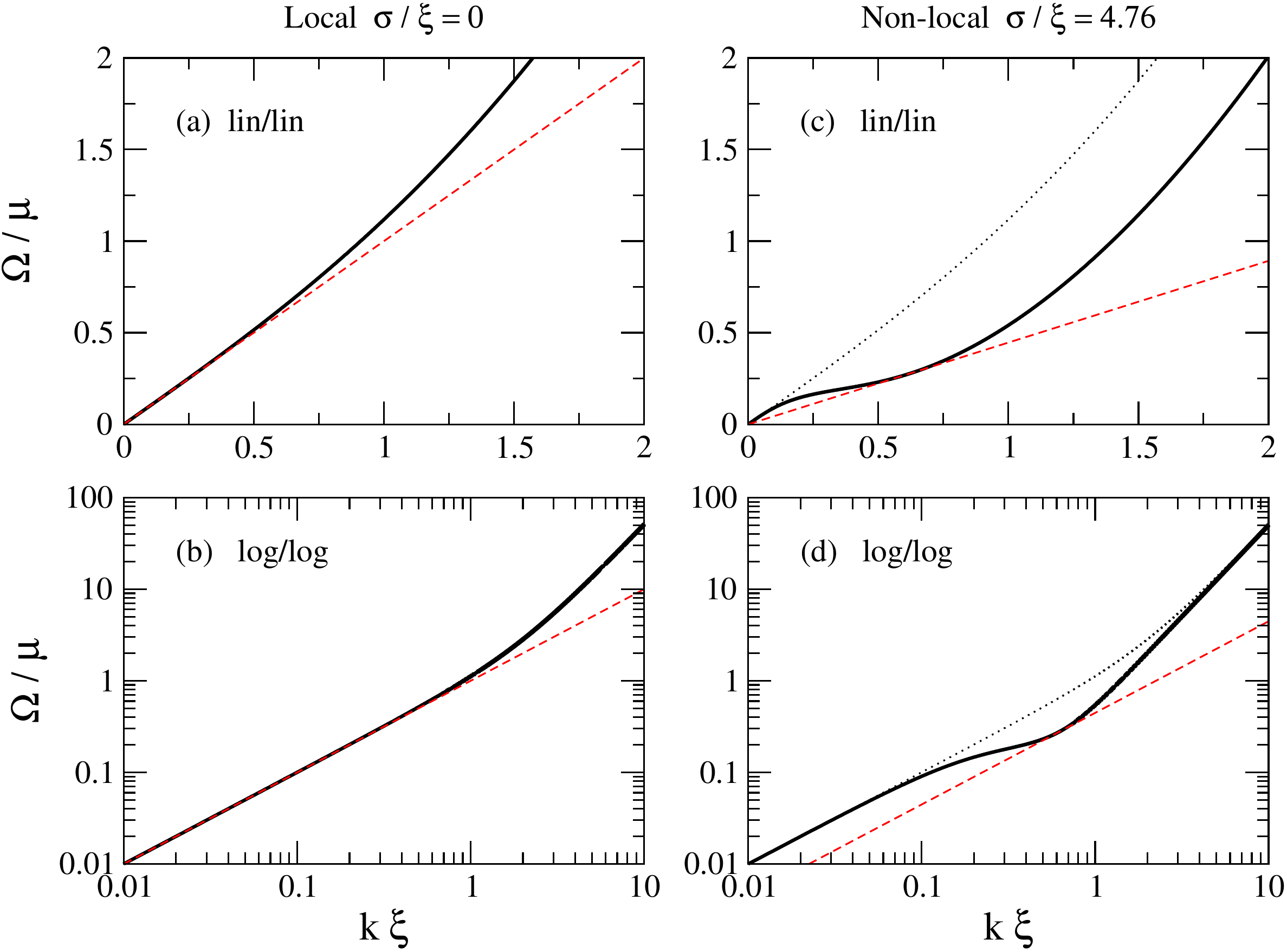}
\caption{Bogoliubov dispersion in a fluid at rest: the left (a,c) panels are for a local fluid $\sigma_L/\xi=0$, the right (b,d) panels are for a nonlocal fluid with a value of $\sigma_L/\xi=4.76$ inspired by the experiment. The dotted curve in the right panels repeats the local fluid dispersion for comparison. The red dashed line is the straight line $\Omega=v_c k$ corresponding to the critical speed $v_c$. The panels in the bottom row show the same curves in log-log scale.}\label{fig:Bogo}
\end{figure}

Examples of such a Bogoliubov dispersion are shown in Fig.\ref{fig:Bogo} both in linear-linear and in log-log scales.
The Landau critical velocity is defined as the minimum of the phase velocity $v_c=\min_k v_{\rm ph}(k)$ where in our case 
\begin{equation}
 v_{\rm ph}(k)=\frac{\Omega(k)}{k}=\left[\frac{\hbar^2k^2}{4m^2}+\frac{g\rho/m}{1+k^2\sigma_L^2}\right]^{1/2}.
\end{equation}

\begin{figure}[htbp]
\includegraphics[width=0.9\columnwidth]{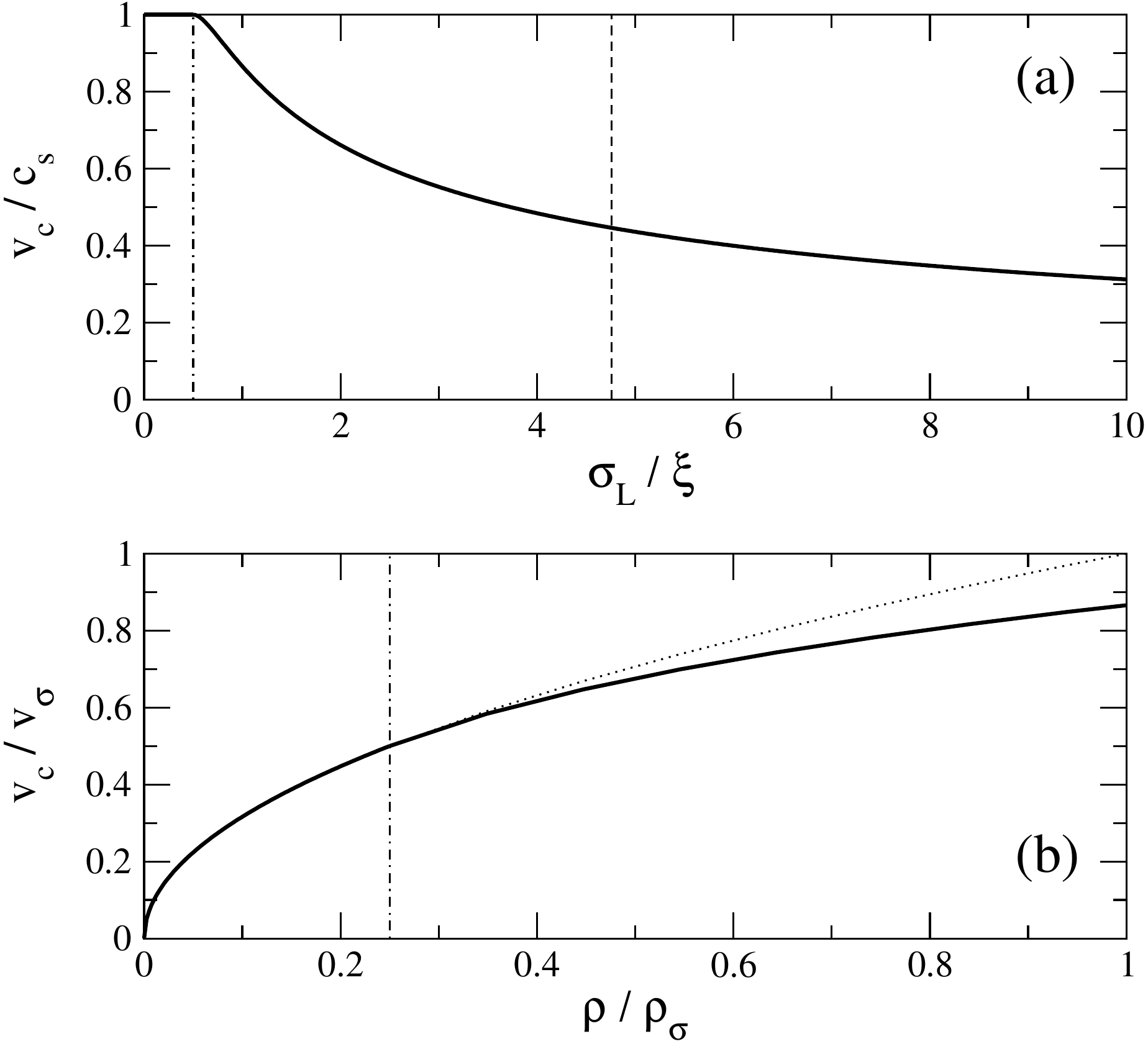}
\caption{Plots of the dependence of $v_c$ on the nonlocal length $\sigma_L$ [upper panel (a)] and on the density $\rho$ [lower panel (b)]. In this latter, the density is normalized to $\rho_\sigma=\hbar^2/mg\sigma_L^2$ and the velocity to $v_\sigma=\hbar/m\sigma_L$. The vertical dot-dashed lines indicate the transition points between the two $\sigma_L \lessgtr \xi/2$
 regimes. The vertical dashed line in the upper panel indicate the experimental conditions. The dotted line in the lower panel shows how $\sqrt{\rho}$ behaviour would extend to the whole domain.} \label{fig:vc}
\end{figure}

Depending on the relative value of the nonlocality length $\sigma_L$ and of the healing length 
\begin{equation}
\xi=\sqrt{\frac{\hbar^2}{g \rho m}}= \sqrt{\frac{n_0}{\Delta n}}\,k_0^{-1},
\end{equation}
two regimes can be identified:
\begin{itemize}
 \item For weak nonlocalities $\sigma_L<\xi/2$ [Fig.\ref{fig:Bogo}(a,b)],
 the phase velocity $v(k)$ is a monotonically growing function of $k$, so the minimum is attained at $k=0$. The critical Landau speed is then the speed of sound, \begin{equation}
v_c=\lim_{k\to 0} \frac{\Omega(k)}{k}=\sqrt{\frac{g\rho}{m}}=\frac{c}{n_0}\sqrt{\frac{\Delta n}{n_0}}=c_s.
\end{equation}

\item For strong nonlocalities $\sigma_L>\xi/2$ [Fig.\ref{fig:Bogo}(c,d)],
 the phase velocity $v(k)$ has a local minimum at 
 \begin{equation}
k_c= \sqrt{\frac{1}{\sigma_L}\left(\frac{2}{\xi}-\frac{1}{\sigma_L}\right)}   
 \end{equation}
where it attains the smaller value 
 \begin{equation}
v_c=\sqrt{\frac{\hbar^2}{m^2\sigma_L}\left[\frac{1}{\xi}-\frac{1}{4\sigma_L}\right]}
\end{equation}
\end{itemize}

Plots of the critical velocity $v_L$ as a function of the nonlocality length and of the fluid density are shown in the two panels of Fig.\ref{fig:vc}. The former clearly shows that the critical velocity is equal to the sound velocity upto $\sigma_L=\xi/2$, then it quickly decays to zero according to the formulas:
\begin{eqnarray}
 v_c/c_s=&1 \;\; &\textrm{for}~\sigma_L<\xi/2 \\
 v_c/c_s=&\sqrt{\frac{\xi}{\sigma_L}-\frac{\xi^2}{4\sigma_L^2}}  \;\; &\textrm{for}~\sigma_L>\xi/2.
 \end{eqnarray}
The latter shows the usual $\sqrt{\rho}$ dependence at low $\rho$, which then transforms into a slower $\rho^{1/4}$ law at high $\rho$ according to the formulas:
\begin{eqnarray}
 v_c=&\sqrt{g\rho/m} \;\; &\textrm{for}~\sigma_L<\xi/2 \\
 v_c=&\frac{\hbar}{m\sigma_L}\sqrt{
\sqrt{\frac{\sigma_L^2 m g \rho}{\hbar^2}}-\frac{1}{4}} \;\; &\textrm{for}~\sigma_L>\xi/2.
 \end{eqnarray}
 Note that the two regimes continuously connect at the transition point $\sigma_L=\xi/2$.
 
For the parameters of the experiment $\Delta n=7.6\times 10^{-6}$ with a background refractive index estimated around $n_0\simeq 1.33$, one has $\sigma_L/\xi\approx 4.76$ which gives a factor $\approx 2$ reduction of the critical velocity $v_c$ below the speed of sound $c_s$.

\end{document}